\documentclass[12pt]{article}

\usepackage{adjustbox, amsmath, amsthm, amsfonts, amssymb, amsbsy, authblk, bm, color, cases, dsfont, float, framed, graphicx, hyperref, enumerate, caption, mathrsfs, makecell, tabularx, multirow, ragged2e, cellspace, booktabs, parskip, scalerel,stmaryrd, tikz}

\usepackage{libertinus}
\usepackage{libertinust1math}
\usepackage[T1]{fontenc}

\usepackage[capitalise]{cleveref}
\usepackage[square,numbers]{natbib}
\usepackage[margin=1in]{geometry}

\newcommand{\fmin}{f_{\mathrm{min}}}
\newcommand{\fmax}{f_{\mathrm{max}}}
\newcommand{\xeq}{x_{\mathrm{eq}}}

\newcommand{\N}{\mathcal{N}}

\begin{document}

\title{How many more is different?}

\author[1,2]{Jacob Calvert\thanks{calvert@gatech.edu}}

\author[3]{Andr{\'e}a W. Richa\thanks{aricha@asu.edu}}

\author[4,2]{Dana Randall\thanks{randall@cc.gatech.edu}}
\affil[1]{Institute for Data Engineering and Science, Georgia Institute of Technology, Atlanta, GA, USA}

\affil[2]{Santa Fe Institute, Santa Fe, NM, USA}

\affil[3]{School of Computing and Augmented Intelligence, Arizona State University, Tempe, AZ, USA}

\affil[4]{School of Computer Science, Georgia Institute of Technology, Atlanta, GA, USA}

\date{}

\maketitle

\begin{abstract}
    From the formation of ice in small clusters of water molecules to the mass raids of army ant colonies, the emergent behavior of collectives depends critically on their size. At the same time, common wisdom holds that such behaviors are robust to the loss of individuals. This tension points to the need for a more systematic study of how number influences collective behavior. We initiate this study by focusing on collective behaviors that change abruptly at certain critical numbers of individuals. We show that a subtle modification of standard bifurcation analysis identifies such critical numbers, including those associated with discreteness- and noise-induced transitions. By treating them as instances of the same phenomenon, we show that critical numbers across physical scales and scientific domains commonly arise from competing feedbacks that scale differently with number. We then use this idea to find overlooked critical numbers in past studies of collective behavior and explore the implications for their conclusions. In particular, we highlight how deterministic approximations of stochastic models can fail near critical numbers. We close by distinguishing these qualitative changes from density-dependent phase transitions and by discussing how our approach could generalize to broader classes of collective behaviors.
\end{abstract}

\section{Introduction}\label{sec1}

The emergent behavior of a collective is vaguely, yet inextricably, linked to the number of its constituents. Consider a colony of ants, the paradigm of collective intelligence. While one hundred army ants exhibit essentially ``aberrant'' behavior, a colony of one million army ants exhibits ``flexible problem solving far exceeding the capacity of the individual'' \cite{franks_army_1989}. More is indeed different, as Philip W.\ Anderson famously observed \cite{anderson_more_1972}. But when---and how abruptly---does a behavior emerge, as a collective becomes more numerous?

\begin{figure}
    \centering
    \includegraphics[width=0.7\textwidth]{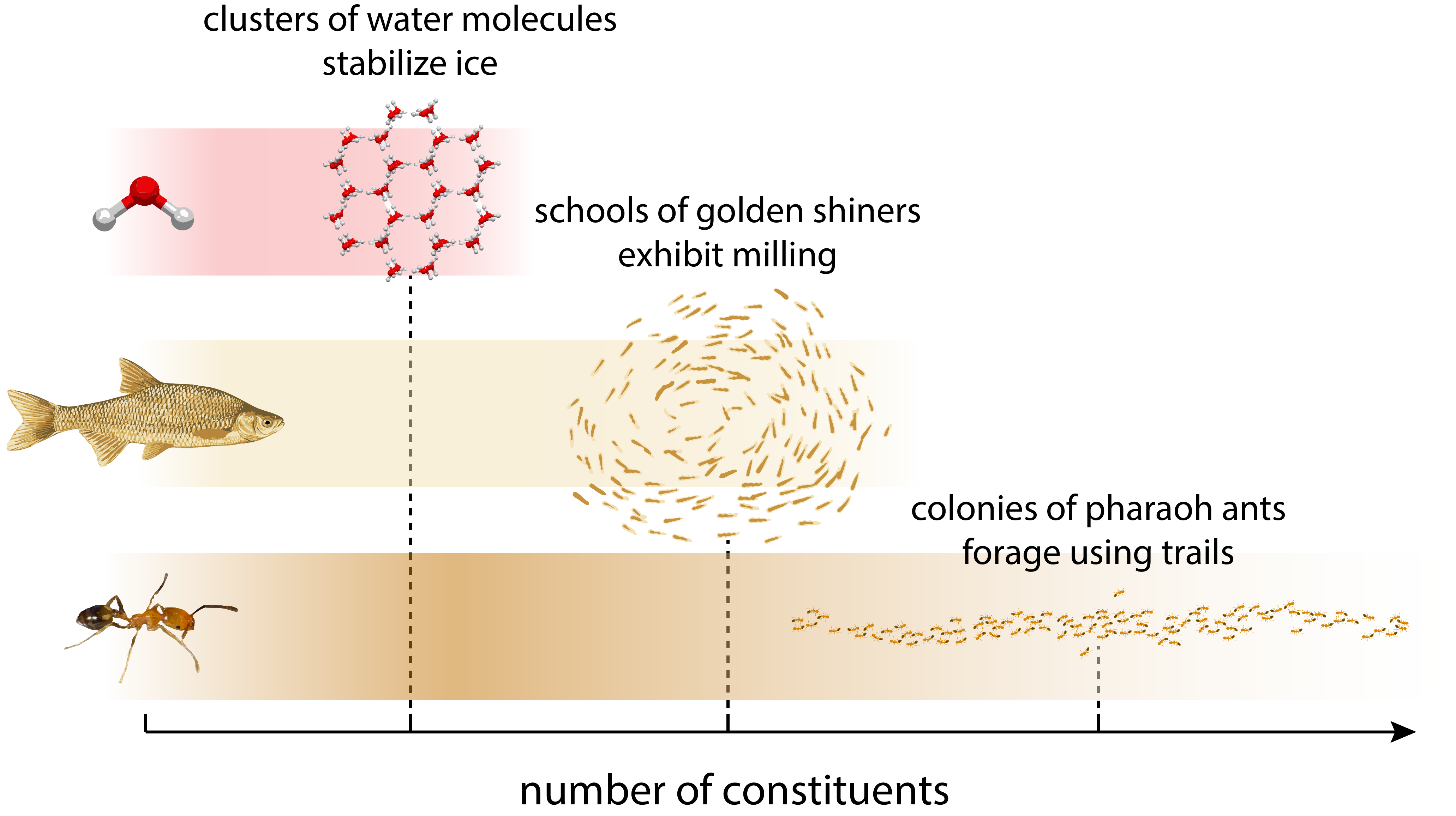}
    \caption{Collectives exhibit qualitatively different behavior as they become more numerous.}
    \label{fig0}
\end{figure}

Experiments have shown that the characteristic behaviors of some animal collectives emerge only when they are sufficiently numerous (\cref{fig0}). For example, \citet{beekman_phase_2001} observed that colonies of $700$ or more pharaoh ants foraged with pheromone trails, while those of $600$ or fewer ants did not. Likewise, \citet{chandra_colony_2021} found that clonal raider ants, which typically forage by group raiding, instead exhibited mass raiding when their colony size was artificially increased. In fact, transitions such as these can occur even when individuals' interactions do not vary with their number. This is what \citet{tunstrom_collective_2013} and \citet{jhawar_noise-induced_2020} concluded from studies of schools of fish, which transitioned from polarized motion to isotropic motion as they became more numerous, even though their local density and social interactions remained similar across group sizes.

Transitions mediated by number are not limited to animal behavior; they occur in systems across physical scales and scientific domains. The phase transition of liquid water to ice is a classic example. Statistical mechanics explains its occurrence in an effectively infinite system of water molecules in terms of a continuous parameter, temperature, which determines the relative strength of molecular interactions and thermal fluctuations. At a fixed temperature below the freezing point, however, the formation of ice is purely a matter of number: A cluster of water molecules needs roughly $90$ or more molecules to sustain ice \cite{moberg_end_2019}. 

These observations suggest that some collective behaviors have specific numbers of constituents, {\em critical numerosities},\footnote{We borrow the word numerosity from \citet{ladyman_what_2020}, who similarly used it to describe the number of constituents of a collective, distinct from size and scale, which could be confused with spatial extent.} at which they abruptly emerge. These qualitative changes in behavior are distinct from phase transitions in statistical mechanics. While it is true that thermodynamic phase transitions technically cannot occur in finite systems, this is not necessarily a problem, as the associated notions of correlation length and susceptibility may still be relevant to finite systems \cite{romanczuk_phase_2022}. The more important difference is that phase transitions in statistical mechanics are usually mediated by the relative strength of interactions, not the number of constituents.

A natural way to explain abrupt changes in behavior with population size is to treat this size as the bifurcation parameter of a dynamical system, with changes in equilibria corresponding to changes in collective behavior. Crucially, however, standard bifurcation analysis fails to detect the transitions in many behaviors, which seem to depend in a fundamental way on the discreteness of the number of individuals or the stochasticity of their interactions. This is why, for example, \citet{jhawar_noise-induced_2020} characterized the schooling of golden shiners as ``noise-induced'' and \citet{saito_theoretical_2015} described an analogous phenomenon in a chemical reaction network as ``discreteness-induced.''

The signature of these phenomena is a qualitative disagreement between two models of the behavior in question \cite{togashi_transitions_2001,biancalani_noise-induced_2014}. The first, which correctly predicts the transition, treats the individuals as discrete and their dynamics as stochastic. Typically, this is a Markov chain or master equation model of the individuals' joint distribution \cite{van_kampen_chapter_2007}. The second, which fails to predict the transition, treats the population as a continuous density with deterministic dynamics. This model is usually an ordinary differential equation (ODE) for the density. Although the stochastic model is generally accepted as a better description of the behavior than the deterministic model \cite{qian_concentration_2002,gillespie_stochastic_2007}, it is more difficult to analyze than the ODE and is often intractable. For this reason, the stochastic model is frequently approximated by the Fokker--Planck equation \cite{gardiner_stochastic_2009} or the system-size expansion \cite{van_kampen_chapter_2007}. But the accuracy of these approximations depends on the number of constituents \cite{grima_how_2011}, which may confound their use for identifying critical numerosities.

\textbf{Main contribution}. We show that critical numerosities can be predicted from the dynamics of constituents by subtly modifying the bifurcation analysis that is standard practice across scientific disciplines \cite{murray_mathematical_2002,sumpter_collective_2010}. Our approach entails analyzing the bifurcations of an ODE which is related to the original model, but not necessarily analogous. While this approach is no more difficult than standard bifurcation analysis, it identifies critical numerosities in models of discrete individuals interacting stochastically as well as models of population densities interacting deterministically. Our approach therefore clarifies that the corresponding transitions in behavior are commonly driven not by noise or by discreteness, but by fewness.

More broadly, the existence of critical numerosities challenges common wisdom about collective behavior. For example, it is often said that the collective behavior of a group of animals is robust to the gain or loss of individuals, because it emerges from the interactions of leaderless individuals \cite{ouellette_goals_2021}. Qualitative changes in the behavior of a collective must therefore come from changes in the nature or strength of individuals' interactions, not their number. This idea, which parallels thinking in statistical mechanics, deeply influences the study of animal groups and other natural collectives, as well as the design of engineered ones. For example, it reinforces the widespread practice of modeling the behaviors of finite collectives as the thermodynamic phases of infinite ones \cite{okubo_dynamical_1986,flierl_individuals_1999,camazine_self-organization_2001,buhl_disorder_2006,bialek_statistical_2012,attanasi_finite-size_2014,ouellette_physics_2022}. It further motivates the design of distributed systems in the likeness of animal groups, with the assumption that they will inherit the same robustness \cite{brambilla_swarm_2013,dorigo_swarm_2014,hamann_swarm_2018}.

The rest of the paper is organized as follows. First, we motivate a precise definition of critical numerosity by demonstrating the failure of standard bifurcation analysis to identify a transition in a model that arises in many fields (\cref{sec: def}). We apply our definition to several models of insect behavior, with comparisons to standard bifurcation analysis, in \cref{sec: examples}. In \cref{sec: influence}, we highlight two ways that hidden critical numerosities affect the interpretation of past studies of collective behavior. The moral of the first example is that ODE approximations of stochastic behavior can fail in the vicinity of a critical numerosity. In other words, in some instances, ODE approximations work well for collectives with few individuals and those with many, but not for ones of an intermediate size! The second example shows how different models of the same experiment implicitly predict critical numerosities that are opposites, in a sense. As a result, one or both of these models must be inaccurate, suggesting new experiments that vary group size. Lastly, in \cref{sec: discussion}, we discuss several important points, including how changes in collective behavior with number differ from density-dependent phase transitions \cite{vicsek_novel_1995,Cates2015}.

\section{Critical numerosities as bifurcations}\label{sec: def}

In this section, we precisely define a unifying notion of critical numerosity for a class of collective behaviors that includes many of the preceding examples. To motivate it, we first show how standard bifurcation analysis fails to identify the bifurcation in a model that arises, in various guises, in chemistry, biology, physics, and economics. We then explain why a subtle modification to the bifurcation analysis, justified by a basic fact about Markov chains, enables the otherwise standard approach to succeed.

\subsection{A model of collective behaviors}\label{subsec:model}

We consider behaviors that can be characterized by a single number $x \in \{0,1,\dots,n\}$, where $n$ is the number of constituents. For example, $x$ could represent the number of ants following a pheromone trail in a colony of size $n$, or the number of people who hold a particular opinion in a population of $n$ individuals. This class of behaviors is broad enough to demonstrate the key ideas about numerosity, and it includes many of the preceding examples. We discuss critical numerosities in higher-dimensional behaviors, like the motion of schooling fish \cite{tunstrom_collective_2013,jhawar_noise-induced_2020}, in a later section.

Concerning the dynamics, we assume that $x$ increases at a rate of $b_n(x)$ and decreases at a rate of $d_n(x)$. To ensure that $x$ remains nonnegative and does not exceed $n$, we assume that $b_n$ and $d_n$ are real-valued functions on the continuous interval $[0,n]$ that satisfy $b_n(n) = d_n(0) = 0$ and are otherwise positive. We further assume that $b_n$ and $d_n$ are differentiable, to simplify the discussion of extrema. These properties are satisfied by all examples we consider.

We assume that, for each integer $n \geq 2$, the dynamics ultimately reaches a steady state described by a probability distribution $\pi_n$ supported on $\{0,1,\dots,n\}$. The probability $\pi_n (x)$ is the long-run fraction of time that the collective spends in state $x$. The peaks of $\pi_n$ correspond to the likeliest or stablest states of the collective; their number reflects qualitative features of the collective behavior \cite{horsthemke_noise-induced_1984,de_palma_bifurcation_1984,vellela_stochastic_2008,mendler_analysis_2018}.

Based on the examples in the introduction, it would be natural to call $n$ a critical numerosity if $\pi_{n-1}$ and $\pi_{n}$ have different numbers of local extrema. (This is closely related to the notion of a phenomenological stochastic bifurcation or ``P-bifurcation'' \cite{horsthemke_noise-induced_1984,arnold_stochastic_1992,arnold_random_1998}.) However, some models could satisfy this definition for some values of $n$ without the shape of $\pi_n$ materially changing, simply because $n-1$ and $n$ have different parities. Rather, we modify this definition of critical numerosity to express essentially the same idea, but without this shortcoming.

\subsection{Bifurcation analysis fails to detect some critical numerosities}\label{subsec:failure}

We first explain in more detail how standard bifurcation analysis fails to detect a critical numerosity in a model of ant foraging. Faced with two, identical paths to a food source, some ant colonies alternately concentrate on one of the paths, instead of using both paths in parallel \cite{beckers_collective_1990,deneubourg_self-organizing_1990}. F{\"o}llmer and Kirman explained this behavior with a simple model, according to which individual ants spontaneously switch between the two paths or are recruited by the ants on the other path \cite{kirman_ants_1993}. Specifically, if $x$ of $n$ ants use the first path, then each such ant spontaneously switches---or is recruited to---the second path at rates of $s$ and $r (n-x)/(n-1)$, respectively. The parameter $r$ is the rate at which each ant meets another ant, while the factor of $(n-x)/(n-1)$ is the probability that an ant foraging on the first path meets one of the $n-x$ ants foraging on the second path. The overall rates at which $x$ increases and decreases are 
\begin{equation}\label{eq: rates of fk model}
    b_n (x) = s (n-x) + \frac{r x (n-x)}{n-1} \quad \text{and} \quad d_n (x) = s x + \frac{r x (n-x)}{n-1}.
\end{equation}

\begin{figure}
    \centering
    \includegraphics[width=0.75\textwidth]{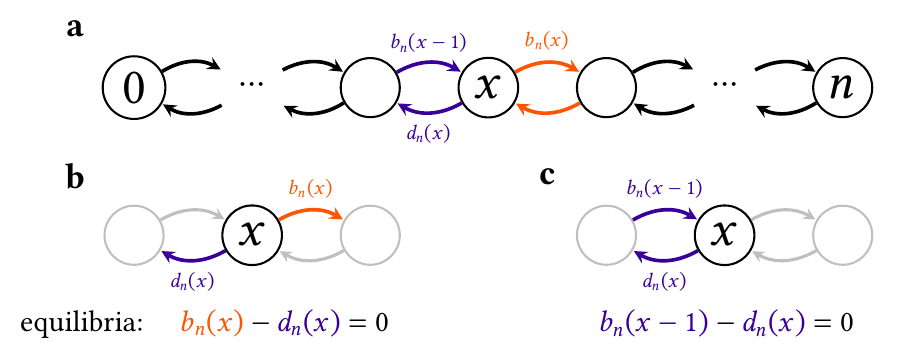}
    \caption{(a) Many models of collective behavior are birth-death Markov chains. (b) A standard bifurcation analysis, which treats the dynamics as deterministic, balances $b_n(x)$ and $d_n(x)$ to identify equilibria. (c) The formula (\cref{eq: birth death}) for the stationary distribution of the Markov chain in (a) indicates that $b_n(x-1)$ and $d_n(x)$ should be balanced instead.}
    \label{fig: birth death chain}
\end{figure}

The F{\"o}llmer--Kirman (FK) model treats the dynamics of the ants as stochastic, with $b_n$ and $d_n$ serving as the rates of a continuous-time Markov chain on the state space $\{0,1,\dots,n\}$ (\cref{fig: birth death chain}a). Simulations of this chain show that as $n$ increases $\pi_n (x)$ goes from having two maxima, at $x \in \{0,n\}$, to having one maximum at $x = n/2$ when $n$ is even, and maxima at $(n\pm 2)/2$ when $n$ is odd (\cref{fig: fk}a,b). In the last case, we view the adjacent maxima as constituting one peak. The transition from two peaks to one, which occurs when $n$ is roughly $r/s$, reflects a qualitative change in collective behavior, from an alternating consensus of ants foraging at one source to a lack of consensus, where ants forage the two sources in parallel (\cref{fig: fk}c).

Now consider what standard bifurcation analysis concludes about the FK model. Pretending for the moment that the dynamics is deterministic and that $x$ takes continuous values, the equilibria of the FK model are the values of $x$ such that
\begin{equation*}
\frac{dx}{dt} = b_n (x) - d_n (x) = s(n-2x) = 0
\end{equation*}
(\cref{fig: birth death chain}b). We would conclude that no bifurcation occurs, because this equation has the same number of stable solutions for every $n$. Specifically, there is a single stable equilibrium at $x = n/2$.

The relevance of this example is not limited to insect behavior. For example, Kirman and others used the FK model to explain herding behavior in financial markets \cite{kirman_ants_1993,alfarano_estimation_2005,alfarano_network_2009,carro_markets_2015}, as well as a variety of other phenomena in behavioral economics \cite{moran_schrodingers_2020}. In this context, the FK model is also known as the noisy voter model \cite{khalil_zealots_2021,caligiuri_noisy_2023}. In biology, the FK model is equivalent to the Moran model of population genetics in the case of two alleles with bidirectional mutation \cite{moran_random_1958}. In chemistry, it arises as the Togashi--Kaneko reaction system with two components \cite{ohkubo_transition_2008,biancalani_noise-induced_2014}. This example inspired the notion of discreteness-induced transitions \cite{togashi_transitions_2001,saito_theoretical_2015}, which was subsequently studied in a series of mathematical works \cite{mcsweeney_stochastically-induced_2014,hoessly_stationary_2019,bibbona_stationary_2020,gallinger_asymmetric_2024}.

\subsection{A modification of bifurcation analysis}

The failure of standard bifurcation theory to capture some transitions is significant in the context of identifying critical numerosities. We now explain how to calculate these critical points in the FK model to demonstrate the subtle but important change to bifurcation analysis. In fact, it is possible to obtain the stationary distribution as well as the full, finite-time distributions of the FK model using probability generating functions \cite{houchmandzadeh_exact_2015,holehouse_exact_2022}. However, as we show, a simpler and more direct method---which extends to more complicated examples---suffices to identify critical numerosities.

The FK model is a {\em birth-death} Markov chain, that is, a chain in which only transitions of the form $x \to x+1$ and $x \to x-1$ are possible. There is an explicit formula for the stationary distributions $\pi_n$ of such chains, in terms of the rates $b_n$ and $d_n$ \cite{pinsky_introduction_2011}:
\begin{equation}\label{eq: birth death}
\pi_n (x) = \pi_n (0) \prod_{y=1}^x \frac{b_n (y-1)}{d_n (y)}, \quad x = 1, 2, \dots, n.
\end{equation}
According to \cref{eq: birth death}, the extrema of $\pi_n$ correspond to values of $x$ for which $b_n (x-1) - d_n (x)$ and $b_n (x) - d_n (x+1)$ have different signs (\cref{fig: birth death chain}c). When $b_n$ and $d_n$ are continuous functions, it is reasonable to approximate these extrema by the values of $x \in [1,n-1]$ for which
\begin{equation}\label{eq: mod equilibria}
b_n (x-1) - d_n (x) = 0.
\end{equation}
This condition is only slightly different than the one that defines equilibria in standard bifurcation analysis (\cref{fig: birth death chain}b). Yet, with this change, the same analysis correctly identifies the critical numerosity in the FK model.

\begin{figure}
    \centering
    \includegraphics[width=0.85\textwidth]{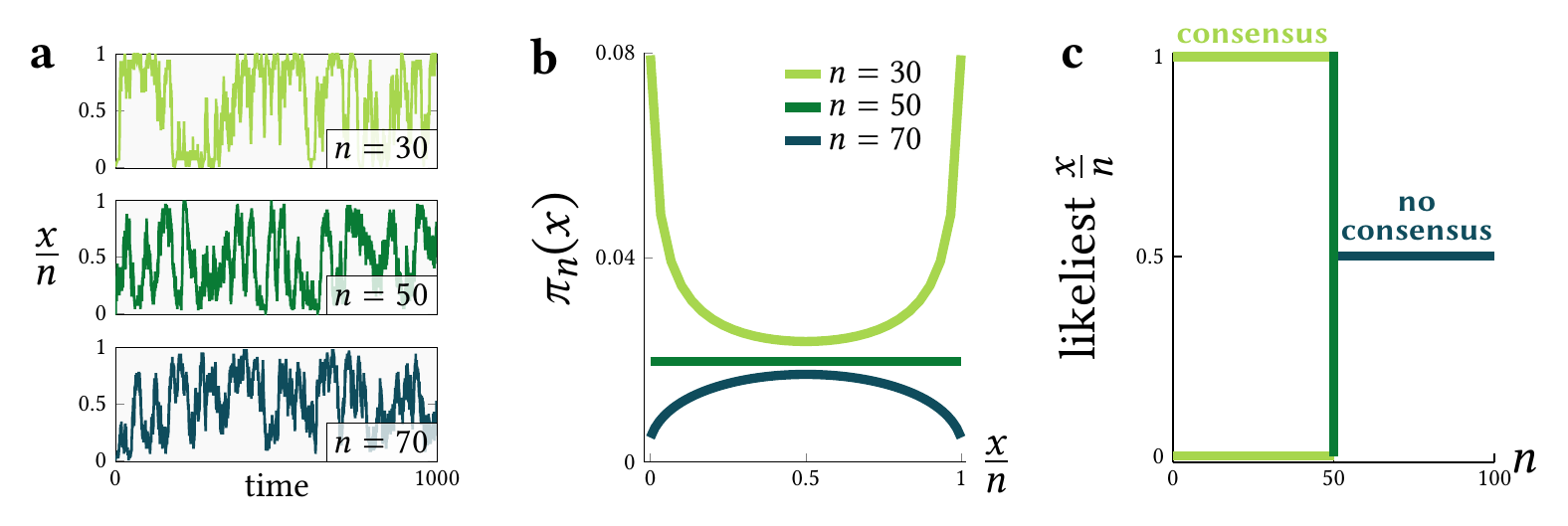}
    \caption{\textbf{Critical numerosity in path selection.} The F{\"o}llmer--Kirman (FK) model describes ants choosing between two identical paths to a food source \cite{kirman_ants_1993}. ({\bf a}) Representative timeseries of the FK model from \cref{eq: rates of fk model} with parameters $r = 1$ and $s = 0.02$.  ({\bf b}) The stationary distribution $\pi_n$ transitions from bimodal to unimodal as the number of individuals $n$ increases from below to above the critical numerosity $n_c \approx 50$. ({\bf c}) This reflects a change in collective behavior, from alternating consensuses to a lack of consensus.}
    \label{fig: fk}
\end{figure}

Substituting the rates of the FK model (\cref{eq: rates of fk model}) into the preceding condition, we find that its extrema approximately correspond to the values of $x \in [1,n-1]$ for which
\[
b_n (x-1) - d_n (x) = \left( s - \frac{r}{n-1} \right) (n+1 - 2x) = 0.
\]
This equation has no stable equilibria when $n$ is at most $r/s + 1$, and one stable equilibrium when $n$ exceeds $r/s + 1$. The FK model therefore has a critical numerosity $n_c$ at roughly $r/s + 1$, which agrees with simulations (\cref{fig: fk}).

\subsection{A precise definition of critical numerosity}

The preceding section motivates the following definition of critical numerosity, which applies to all the models of collective behavior described in \cref{subsec:model}. Given rates $b_n$ and $d_n$, for every integer $n \geq 2$, we denote by $P(n)$ the number of stable equilibria, where equilibria are values of $x \in [1,n-1]$ that are solutions to $b_n (x-1) - d_n (x) = 0$. We then define the set $\N_c$ of critical numerosities to consist of values of $n$ at which the number of stable equilibria changes:
\begin{equation}\label{eq: def of critical numerosity}
    \N_c = \left\{n \geq 2: P(n) \neq P(n-1) \right\}.
\end{equation}

The virtue of this definition is that it involves only a minor change to the definition of equilibria. It otherwise proceeds according to a standard bifurcation analysis. We emphasize that \cref{eq: def of critical numerosity} treats the critical numerosities of a stochastic model as the bifurcations of a related, but not necessarily analogous model, in which the dynamics are deterministic and $x$ takes continuous values. This observation suggests that the transitions associated with critical numerosities are not necessarily noise- or discreteness-induced. As the FK model and the following examples demonstrate, what these transitions have in common is that they are induced by fewness. 

\section{Critical numerosities in models of living systems}\label{sec: examples}

To further highlight the relevance of critical numerosities and demonstrate the use of \cref{eq: def of critical numerosity}, we discuss three additional models of insect collective behavior. The naive use of standard bifurcation analysis identifies the critical numerosities in the first two models, but it fails in the third. For ease of comparison, we use the same notation for the parameters of the models as their papers of origin. 

\subsection{Trail formation}\label{subsec: trail formation}

\begin{figure}
    \centering
    \includegraphics[width=0.85\textwidth]{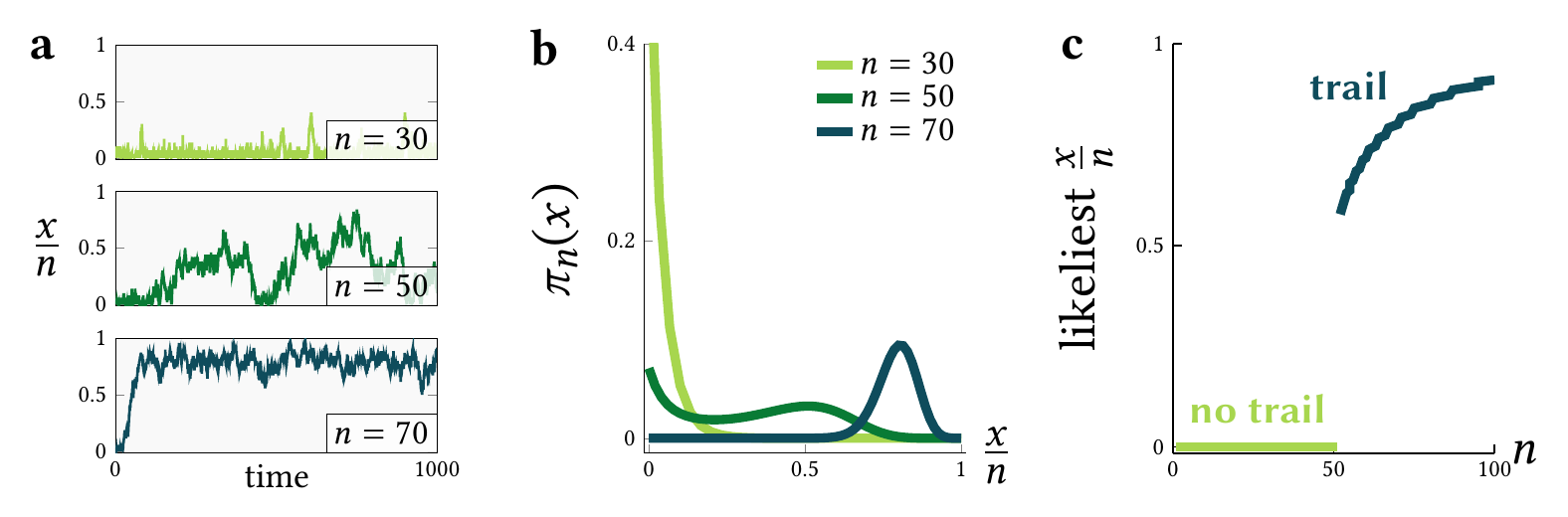}
    \caption{\textbf{Critical numerosity in trail formation.} The Beekman--Ratnieks--Sumpter (BRS) trail formation model describes a colony of ants attempting to forage using a pheromone trail \cite{beekman_phase_2001}. ({\bf a}) Representative timeseries of the BRS model with parameters $q = 0.03$, $r = 0.002$, and $s = 2$. ({\bf b}) As $n$ increases beyond the critical numerosity $n_c \approx 50$, the stationary distribution forms a new peak at a positive value of $x$. ({\bf c}) This reflects a change in collective behavior from the absence of a trail to a trail that the majority of ants follow.}
    \label{fig: brs}
\end{figure}

\citet{beekman_phase_2001} demonstrated that, under certain experimental conditions, pharaoh ants forage using a pheromone trail if and only if the colony is sufficiently numerous. They explained this observation using a model according to which each ant either follows the trail or does not, and the state of the model is determined by the number $x$ of ants following the trail. Each of the $n-x$ ants not on the trail independently finds it at a rate $q$ and is attracted to it at a rate of $r x$, while each ant on the trail loses it at a rate of $s/(s+x)$. The overall rates at which the number of ants on the trail increases and decreases are
\[
b_n(x) = (q + r x) (n-x) \quad \text{and} \quad d_n(x) = \frac{sx}{s+x}.
\]
For simplicity, because $sx/(s+x)$ is approximately $s$ when $x$ is large relative to $s$, we replace $d_n$ with the constant $s$. \cref{fig: brs}a shows simulations of the model with this simplification. Following \cref{sec: def}, we find that the net rate from $x-1$ to $x$ is a quadratic polynomial in $x$:
\[
b_n(x-1) - d_n(x) = -r x^2 + \left( r(n+2) - q \right) x + \left( (q-r)(n+1) - s \right).
\]
The corresponding equilibria are
\[
    \frac12 \left(n + 2 - q/r\right) \pm \frac12 \sqrt{\left(n+q/r\right)^2 - 4s/r}.
\]
For further simplicity, we assume that $q/r \geq 3$, which suffices to ensure that these values of $x$ lie in $[1,n-1]$. If $(n+q/r)^2$ is less than $4s/r$, then there are no extrema, because we only consider real values of $x$. But there are two extrema when $(n+q/r)^2$ is larger, the greater of these extrema corresponding to a local maximum (\cref{fig: brs}b). This implies that there is a critical numerosity at roughly $2\sqrt{s/r} - q/r$. Below it, the ants are unable to form a trail, while they can and do form a trail above it (\cref{fig: brs}c).

\subsection{Shelter selection}

The German cockroach aggregates with conspecifics during periods of rest in shelters with favorable physical characteristics \cite{rust_understanding_1995,appel_harborage_1996}. \citet{ame_cockroach_2004} conducted experiments in which various numbers of cockroaches were offered two identical shelters. On the basis of these experiments, they fitted probabilities with which a cockroach would leave a shelter, based on the number of conspecifics at the shelter. They found that, when there were $n$ cockroaches in total, the rates at which the number $x$ of cockroaches at the first site increased and decreased were
\begin{equation}\label{eq: rates of ame model}
    b_n (x) = \frac{\theta(n-x)}{k + (n-x)^2} \quad \text{and} \quad d_n (x) = \frac{\theta x}{k + x^2},
\end{equation}
for positive numbers $\theta$ and $k$.

Some algebra shows that the value of $\theta$ is irrelevant to the modified equilibria of \cref{eq: rates of ame model}, which are
\[
\frac{n+1}{2}, \quad \frac{n+1}{2} \left(1 \pm \sqrt{1 - 4k / (n+1)^2} \right).
\]
It is easy to check that there is a critical numerosity $n_c$ at roughly $2\sqrt{k}$, when the term in the square root becomes real. Below $n_c$, the cockroaches split evenly between the two shelters. Above it, they disproportionately aggregate at one of the shelters.

\subsection{Task allocation}\label{subsec: task switching}

\citet{pacala_effects_1996} modeled the allocation of $n$ social insects to two tasks, labeled $i \in \{1,2\}$. According to their model, individuals succeed at task $i$ with a probability of $s_i$. Unsuccessful individuals spontaneously encounter the stimulus for the other task at a rate of $f_i$ or are recruited to it at a rate of $s(1-s) n^{-\alpha}$, for a number $\alpha \in [0,1]$. The factor $n^{-\alpha}$ corresponds to the per capita rate of social interactions, which interpolates between the extremes of a constant rate when $\alpha = 0$ and a rate that grows linearly with $n$ when $\alpha = 1$. For tasks that are equally difficult, i.e., when $s_1 = s_2 = s$, the number $x$ of individuals allocated to the first task increases and decreases at rates of
\begin{equation}\label{eq: rates of pacala2 model}
    b_n (x) = f_1 (1-s) (n-x) + \frac{s (1-s) x (n-x)}{n^\alpha} \quad \text{and} \quad d_n (x) = f_2 (1-s) x + \frac{s (1-s) x (n-x)}{n^\alpha}.
\end{equation}
In the special case where $f_1 = f_2$ and $\alpha = 1$, the rates are essentially a scaled version of the FK model's rates.

\begin{figure}
    \centering
    \includegraphics[width=0.85\textwidth]{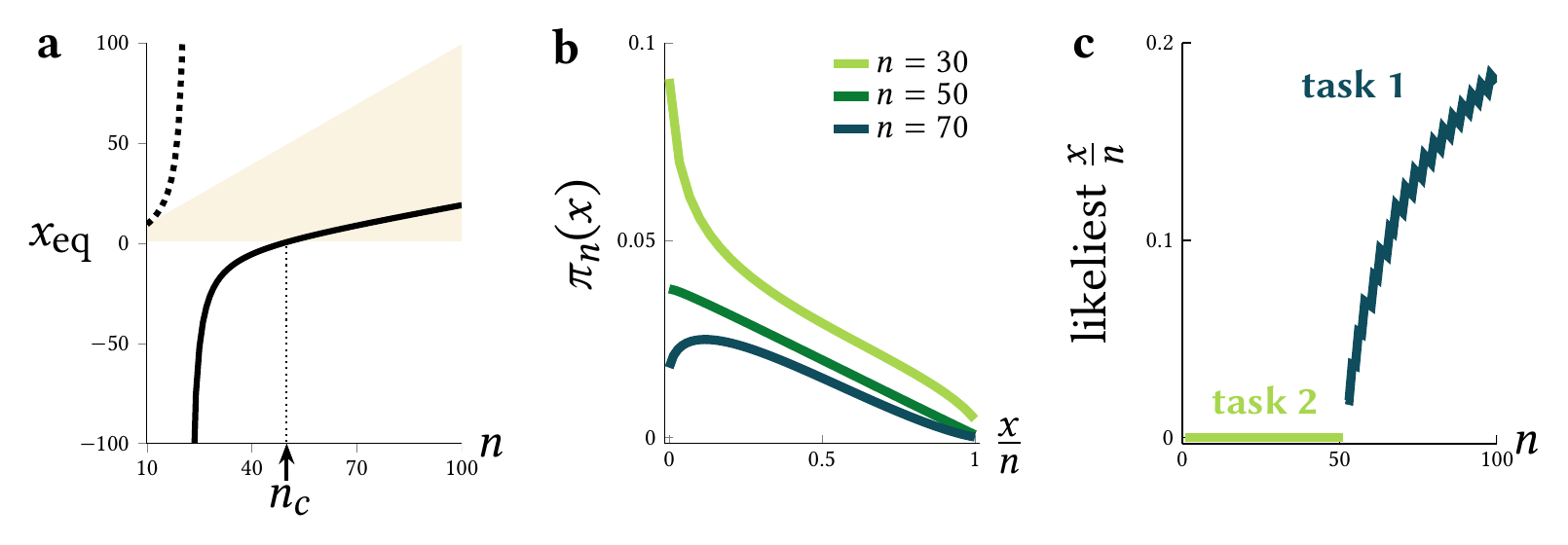}
    \caption{\textbf{Critical numerosity in task allocation.} Pacala et al.\ model the allocation of social insects to one of two tasks. ({\bf a}) The model has one modified equilibrium, $x_{\mathrm{eq}}$, which is unstable (dashed line) and lies outside the interval $[1,n-1]$ (shaded region) when $n$ is small. As $n$ increases, $x_{\mathrm{eq}}$ becomes stable (solid line) and enters the interval at $n_c = 50$. ({\bf b}) The entry of $x_{\mathrm{eq}}$ into the interval $[1,n-1]$ produces a peak of $\pi_n (x)$ at a strictly positive value of $x$. ({\bf c}) This peak reflects a change in collective behavior from ants concentrating on the second task to increasingly working on the first.}
    \label{fig: pacala2}
\end{figure}

The rates in \cref{eq: rates of pacala2 model} have one modified equilibrium, which is given by
\[
\xeq = \frac{f_1 - s n^{-\alpha}}{f_1 + f_2 - 2sn^{-\alpha}} (n+1).
\]
While only one value of $x$ is an equilibrium, there is a critical numerosity near $(s/\min\{f_1,f_2\})^{1/\alpha}$ (\cref{fig: pacala2}a). As we detail in \cref{app: task switching}, $\xeq$ is unstable for sufficiently small values of $n$. But, as $n$ increases, $\xeq$ exits the domain $[1,n-1]$, stabilizes, and then re-enters the domain, which produces the critical numerosity (\cref{fig: pacala2}b). In terms of behavior, the model predicts that, as $n$ increases, the collective initially abandons the task with the stimulus it encounters less frequently to concentrate on the other task (\cref{fig: pacala2}c). Sufficiently large collectives allocate individuals to both tasks, albeit unevenly.

Unlike the preceding examples, standard bifurcation analysis fails to identify this critical numerosity. Indeed, treating $b_n$ and $d_n$ as the rates of an ODE model, the corresponding equilibria solve
\[
\frac{dx}{dt} = b_n (x) - d_n (x) = f_1 (1-s) (n-x) - f_2 (1-s) x = 0.
\]
There is one equilibrium for every $n$, at $f_1 n / (f_1 + f_2)$.

\section{Unseen influence of numerosity on studies of collectives}\label{sec: influence}

Critical numerosities arise in models of collective behavior across scientific domains. However, their influence on the conclusions of the associated studies is typically underappreciated. In this section, we highlight two striking examples. The first concerns the perennial challenge of assessing the accuracy with which an ODE model approximates a stochastic one. The second concerns different models of the same collective animal behavior that predict opposing critical numerosities, implying that at least one of these models is inaccurate.

\subsection{ODE approximations can fail in the vicinity of a critical numerosity}

Many stochastic models, including models of chemical reaction networks, epidemics, and populations, converge to their deterministic analogues, in a sense, under an appropriate scaling limit \cite{kurtz_solutions_1970,kurtz_limit_1971,kurtz_strong_1977}. For example, if the rates of a chemical reaction network are scaled by volume according to mass action kinetics, then the master equation converges to its ODE counterpart in the limit of infinite volume \cite{kurtz_relationship_1972}. When the concentrations of the chemical species are fixed, the limit of infinite volume corresponds to a limit of infinite counts of each molecule. This result is widely cited because it justifies the use of ODE models in the place of master equations---not just in the modeling of chemical reaction networks---but wherever suitable master equations arise.

Although this convergence result is well known, it has two subtle aspects that lead to surprising conclusions in studies of collective behavior. First, the convergence it describes applies to the solutions of the master equation and ODE models on {\em finite} intervals of time. In particular, it does not preclude the disagreement of the {\em steady-state} solutions to the master equation and corresponding ODE model. This fact underlies Keizer's paradox, the observation that stochastic and deterministic models of populations in which extinction is possible but rare predict different steady states, even in the limit of infinite volume or number \cite{keizer_statistical_1987,vellela_quasistationary_2007}. Second, even when the steady-state solutions of the two models do agree in the limit, the convergence result does not indicate how numerous the constituents must be before the two models are ``close.'' Moreover, the disagreement between the models does not necessarily decrease monotonically as volume or number grow. As the next example demonstrates, ODE approximations to stochastic models of collective behavior can fail near a critical numerosity.

\citet{pacala_effects_1996} primarily used ODE models to study of the effect of animal social group size on the allocation of individuals to tasks. (This includes the model we highlighted in \cref{subsec: task switching}, as well as a second model that we introduce below.) To support the relevance of their conclusions to the small social groups that occur in nature, they compared the mean number of individuals allocated to each task under the ODE models to simulations of their stochastic analogues. They found that, remarkably, the predictions of the deterministic models agreed with the stochastic models, even for groups with as few as ten individuals. On the basis of this observation, they concluded that, at least in some cases, the behavior predicted by the ODE models occurs in small groups of individuals.

While this is true, the agreement between the ODE models and their stochastic analogues is highly sensitive to the model parameters, in a way we now explain. In fact, the stochastic models have critical numerosities immediately above which their means can vastly differ from those of the deterministic models. For the parameters that Pacala et al.\ simulated, the critical numerosity is $3$, which is why they observed increasing agreement as group size ranged from tens to thousands. However, for slightly different values of the parameters, the critical numerosity appears in the range of group sizes that Pacala et al.\ simulated.

The model concerns $n$ individuals, $x$ of which are active and $n-x$ of which are inactive. Each inactive individual spontaneously encounters the stimulus for activity at a rate of $f > 0$ or is recruited to the task by active individuals. Recruitment requires that an inactive individual encounters an active one, which depends on the per capita rate of interactions $a > 0$, as well as the probability $s \in (0,1)$ that the newly recruited individual succeeds at the task. Active individuals who fail at the task instead become inactive at a rate of $q > 0$. In summary, the model has rates of  
\begin{equation}\label{eq: pacala1 rates}
b_n(x) = (f + as x) (n-x) \quad \text{and} \quad d_n(x) = q(1-s)x.
\end{equation}

\begin{figure}
    \centering
    \includegraphics[width=0.8\textwidth]{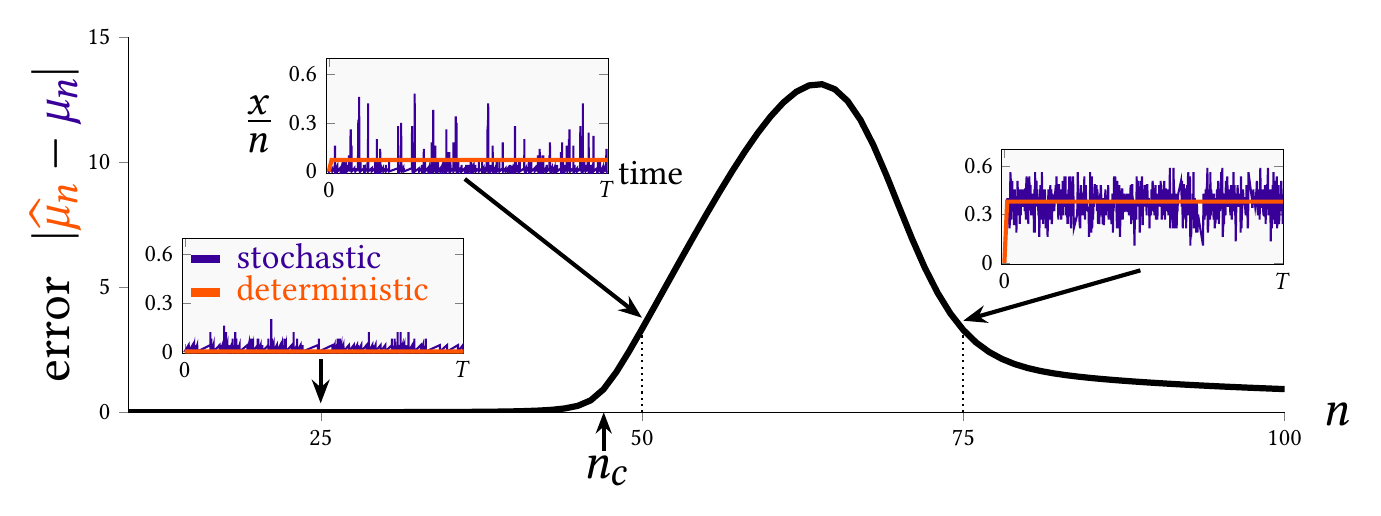}
    \caption{\textbf{ODE approximation of a stochastic model fails near a critical numerosity.} The mean $\mu_n$ of the stochastic model of \citet{pacala_effects_1996} (blue) is closely approximated by the equilibrium $\widehat{\mu}_n$ of the corresponding deterministic model (red), except when $n$ is close to the critical numerosity $n_c = 47$. The inset timeseries of the fraction $x/n$ of active individuals show that the collective behavior changes from exclusively inactive individuals below $n_c$ to roughly one third of individuals being active above it. The plots use the parameters $a = 0.01$, $f = 5 \times 10^{-5}$, $q = 0.2$, and $s = 0.3$, and  in \cref{eq: pacala1 rates}, and the timeseries last $T = 5/f$ units of time.}
    \label{fig: ode fails}
\end{figure}

A calculation like the one in \cref{subsec: trail formation} shows that there is a critical numerosity at roughly $(f+q(1-s))/as$. \citet{pacala_effects_1996} simulated the stochastic model for the parameters $a = 0.1$, $f=0.005$, $q = 0.1$, and $s = 0.3$. In this case, the critical numerosity is $n_c = 3$ and the error, or absolute difference between the deterministic mean $\widehat\mu_n$ and the mean $\mu_n$ of the stochastic model, is uniformly close to $0$ across a wide range of group sizes. However, for somewhat different values of the parameters, the critical numerosity appears in this range, and the error spikes above it (\cref{fig: ode fails}). In particular, the error does not decrease monotonically as group size increases.

\subsection{Critical numerosities differentiate models for the same phenomenon}

\citet{pasteels_self-organization_1987} conducted an experiment in which a colony of pavement ants foraged two identical food sources. Strikingly, they observed that the colony eventually foraged the two sources alternately, as opposed to foraging them in parallel. At least three models have been proposed to explain why alternate foraging arises. The first model, offered by Pasteels et al.\ and further studied in \cite{beckers_collective_1990}, is a relatively complicated system of coupled ODEs that accounts for the ants at each source, lost ants, and ants at the nest \cite{pasteels_self-organization_1987}. The second, introduced by \citet{deneubourg_self-organizing_1990}, models an analogous (but more easily controlled) experiment in which ants are offered two identical paths to the same food source. Each ant chooses a path with a probability depending on the pheromone concentration on the path, which in turn depends on the number of ants on the path. We detail these models in \cref{app: sim parameters}. The third is the FK model, which we discussed in \cref{subsec:failure}.

\begin{figure}
    \centering
    \includegraphics[width=0.8\textwidth]{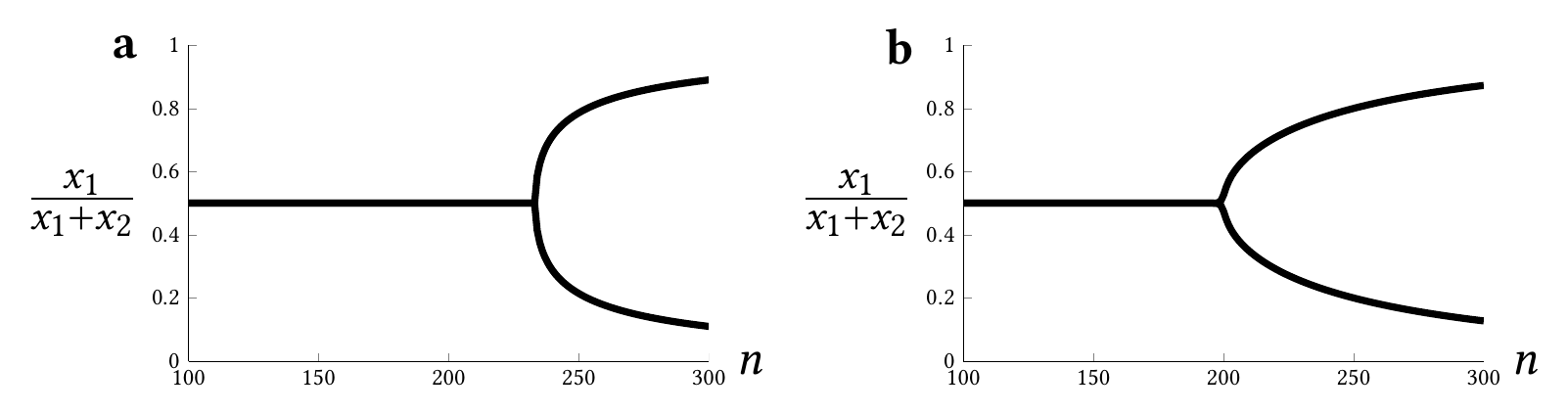}
    \caption{\textbf{Critical numerosities in competing models of ant foraging.} The models of ({\bf a}) \citet{pasteels_self-organization_1987} and ({\bf b}) \citet{deneubourg_self-organizing_1990} both predict that small colonies forage two identical sources in parallel, while large colonies forage them in alternation. The FK model predicts the opposite (\cref{fig: fk}c). Here, $x_1$ and $x_2$ denote the equilibrium numbers of ants in a colony of size $n$ foraging the first and second sources.}
    \label{fig: deneubourg}
\end{figure}

In fact, all three models predict that foraging has a critical numerosity. But the behaviors they predict above and below it differ. While the FK model predicts that the colony transitions from alternate to parallel foraging as it becomes more numerous (\cref{fig: fk}c), the other two models predict the opposite (\cref{fig: deneubourg})! To our knowledge, no experiment has confirmed which (if any) of these transitions occurs.\footnote{While both \citet{pasteels_self-organization_1987} and \citet{deneubourg_self-organizing_1990} mention bifurcations in foraging behavior, these refer to bifurcations in {\em time}---transient foraging in parallel giving way to sustained foraging in alternation. In fact, neither paper describes observing sustained foraging in parallel, although their models predict it for certain colony sizes.} Nevertheless, the foraging example highlights another way that critical numerosities arise in models of collective behavior and motivate experiments for their study.

\section{Discussion}\label{sec: discussion} 

There is a need for systematic study of the dependence of collective behavior on number. This need is evidenced by the incompatibility of experiments demonstrating collective behaviors that depend sensitively on number \cite{beekman_phase_2001,tunstrom_collective_2013,jhawar_noise-induced_2020,ko_small_2022} and common wisdom stating that such behavior is robust to the loss of individuals \cite{sahin_swarm_2005,brambilla_swarm_2013,ouellette_goals_2021}. It is further evidenced by unseen critical numerosities in models of collective behavior \cite{pasteels_self-organization_1987,deneubourg_self-organizing_1990,pacala_effects_1996}, which can affect the conclusions or interpretations of the corresponding studies. For example, ODE approximations of stochastic models can be highly inaccurate near a critical numerosity, resulting in counterintuitively non-monotonic error as the number of individuals grows (\cref{fig: ode fails}). Moreover, different models of the same collective behavior can be differentiated by their behavior in the vicinity of a critical numerosity (\cref{fig: fk,fig: deneubourg}).

We sought to initiate this study by defining a unifying framework for critical numerosities by treating their existence in models across physical scales and scientific domains as part of the same phenomenon. We defined critical numerosities as numbers of individuals at which the distribution over collective behavior gains or loses a peak, corresponding to a mode of collective behavior. We then reconciled the critical numerosities that arise as standard bifurcations (\cref{fig: brs}) from those that appear to inherently depend on the discreteness of individuals or the stochasticity of their interactions (\cref{fig: fk,fig: pacala2}). We did so by showing that the critical numerosities in all these examples can be identified through a subtle change to standard bifurcation analysis (\cref{fig: birth death chain}). As a consequence, we can view these critical numerosities as having a common cause: competing feedbacks that scale differently with number.

In some cases, it is difficult to disentangle number and density. For example, the experiments of \citet{beekman_phase_2001} were conducted in an arena of fixed size, hence the global density of the colony grew in step with its size. In other cases, however, it is clear that number---and not density---is responsible for the change in collective behavior. This is true of collectives that regulate the local density of individuals or the per capita rate of social interactions, like midges \cite{Kelley2013,puckett_determining_2014}, fish \cite{tunstrom_collective_2013,jhawar_noise-induced_2020}, and various species of ants \cite{Gordon1993,pacala_effects_1996}. Of course, global density matters in the sense that, if individuals were so separated as to not interact at all, then collective behavior would not possible. But this is not an explanation of the behavioral transitions that localized collectives exhibit.

The particular value of a critical numerosity can depend on many factors. For example, the value of $n_c$ in the trail formation model depends on parameters that encapsulate the volatility of the pheromone and the geometry of the arena in which the colony forages \cite{beekman_phase_2001}. Another example comes from \citet{ko_small_2022}, who explained why $10$ or more fire ants stably form a raft, while fewer do not, in terms of the competition between surface tension and the Cheerios effect. This competition, in turn, relates to the ants' physical characteristics. Critical numerosities further depend on the precise definition of the behavior in question, and reasonable alternatives can produce different values of $n_c$. What is more fundamental---and what has the potential to persist across related experimental conditions and definitions of behavior---is the abruptness of the transition associated with a critical numerosity.

A natural direction for future work is to develop an analogous theory for behaviors described by the modes of higher-dimensional distributions, such as the motion of fish schools \cite{tunstrom_collective_2013,jhawar_noise-induced_2020}. In the same way that our approach for one-dimensional behaviors uses an exact formula for the stationary distribution of birth-death Markov chains \eqref{eq: birth death}, it may be possible to use exact formulas for the stationary distributions of other classes of Markov chains, like complex-balanced chemical reaction networks, to identify critical numerosities in higher-dimensional behaviors \cite{anderson_product-form_2010,hoessly_stationary_2019}.\footnote{The relevance of these chemical reaction networks is not limited to chemistry; like the FK model, they have many equivalents across domains.} A second possible approach, which avoids exact formulas, involves deriving a Fokker--Planck (FP) equation for the distribution over collective states and subsequently analyzing the fixed points of its convection term \cite{qian_concentration_2002,mendler_analysis_2018,mendler_predicting_2020,becker_relation_2020}. The strength of this approach is that it does not require solving the FP equation. One possible weakness is that the FP entails an approximation, the quality of which depends on the number of individuals. Approaches based on van Kampen's system-size expansion \cite{van_kampen_chapter_2007}, like those in \cite{biancalani_noise-induced_2014} and \cite{PhysRevE.92.052708}, have the same issue. Nonetheless, these approximations may be accurate enough in practice to identify the critical numerosities of many collective behaviors \cite{grima_how_2011}.

\subsection*{Acknowledgments}

This work was supported by U.S.\ Army Research Office award MURI W911NF-19-1-0233.  DR and JC were also supported in part by 
the National Science Foundation award  CCF-2106687 and
AR was supported in part by the National Science Foundation a
ward CCF-2106917.

\appendix

\section{Additional model details}\label{app: calc details}

\subsection{Task allocation}\label{app: task switching}

In \cref{subsec: task switching}, we identified a modified equilibrium in the task switching model of \citet{pacala_effects_1996} at
\[
\xeq = \frac{f_1 - sn^{-\alpha}}{f_1+f_2-2sn^{-\alpha}}(n+1).
\]
Note that $\xeq$ differs from the equilibrium $f_1 n/(f_1+f_2)$ of the corresponding ODE model. To identify the critical numerosities associated with $\xeq$, we check whether it is stable and whether it lies in $[1,n-1]$, as $n$ varies. Concerning the former, we have the sufficient conditions
\[
\xeq \in [1,n-1] \quad \Longleftarrow \quad
\begin{cases}
    n \leq (s/\fmax)^{1/\alpha} & f_1+f_2 \leq 2sn^{-\alpha},\\
    n \geq (s/\fmin)^{1/\alpha} & f_1+f_2 > 2sn^{-\alpha},
\end{cases}
\]
in terms of the minimum $\fmin$ and maximum $\fmax$ of $\{f_1,f_2\}$. In fact, these conditions are approximately necessary, and we will treat them as such for simplicity. Concerning the latter, the stability of $\xeq$ requires that
\[
\frac{d}{dx} \left( b_n (x-1) - d_n (x) \right) \, \, \Bigg|_{\xeq} < 0 \iff \frac{2s(1-s)}{n^{\alpha}} - (f_1+f_2)(1-s) < 0 \iff n > \left( \frac{2s}{f_1+f_2} \right)^{1/\alpha}.
\]
When $f_1+f_2 \leq 2sn^{-\alpha}$, the domain and stability conditions are incompatible because
\[
\left( \frac{2s}{f_1+f_2} \right)^{1/\alpha} \geq \left( \frac{s}{\fmax} \right)^{1/\alpha}.
\]
In contrast, when $f_1+f_2 > 2sn^{-\alpha}$, the domain condition subsumes the stability one. We conclude that there is a critical numerosity at roughly $(s/\fmin)^{1/\alpha}$.

\subsection{Simulation parameters}\label{app: sim parameters}

In this section, we detail the models shown in \cref{fig: deneubourg}. The first model, from \citet{pasteels_self-organization_1987}, is a system of coupled ODEs:
\begin{align*}
    \frac{dx_1}{dt} &= a x_1 f_1 (n-x_1-x_2-y) - bx_1 + cy\\
    \frac{dx_2}{dt} &= a x_2 f_2 (n-x_1-x_2-y) - bx_2 + cy\\
    \frac{dy}{dt} &= a \big(x_1 (1-f_1) + x_2 (1-f_2)\big) (n-x_1-x_2-y) - py - 2cy.
\end{align*}
Here, $n$ is the number of workers able to forage; $x_1$ and $x_2$ denote the number of ants foraging the first and second sources at time $t$; and $y$ denotes the number of ``lost'' ants. The quantities $f_1$ and $f_2$ denote how efficiently ants follow a trail, with $f_i = x_i/(g+x_i)$ for $i \in \{1,2\}$. The quantities $a$, $b$, $c$, $g$, and $p$ are constants representing various rates and probabilities. Following \cite{pasteels_self-organization_1987}, we use $a = 0.001$, $b=0.1$, $c=0.018$, $g=25$, and $p=0.033$. We defer to \citet{pasteels_self-organization_1987} for their dimensions and interpretation. \cref{fig: deneubourg}a plots the ratio $x_1/(x_1+x_2)$ for the equilibrium values of $x_1$ and $x_2$, as a function of the number of workers $n$.

The second model, shown in \cref{fig: deneubourg}b, is that of \citet{deneubourg_self-organizing_1990}. According to their model, the probability $p_i (t)$ with which an ant follows path $i \in \{1,2\}$ at time $t$ is a nonlinear function of the concentration of pheromone $C_i (t)$ on the path:
\begin{equation*}
p_i (t) = \frac{(k+C_i(t))^\alpha}{(k+C_i(t))^\alpha + (k+C_{\neq i}(t))^\alpha}.
\end{equation*}
As Nicolis and and Deneubourg explain \cite{nicolis_emerging_1999}, the concentration of pheromone $C_i(t)$ is proportional to the number of ants on path $i$. As $C_i(t)$ increases past $k$, $p_i(t)$ increases to $1$ with a steepness determined by $\alpha$. The values $k = 6$ and $\alpha = 2$ were fitted from experiments with black garden ants \cite{beckers_trails_1992,beckers_modulation_1993}. The Deneubourg et al.\ model incorporates $p_i(t)$ into a pair of coupled ODEs that describe the time evolution of the pheromone concentration, hence the number of ants, on each path:
\[
\frac{d C_i (t)}{dt} = \phi q p_i(t) - \nu C_i.
\]
The parameter $\phi$ is the rate at which ants leave the nest, which scales with the number of ants in the colony \cite{nicolis_emerging_1999}, $q$ is the rate at which ants deposit pheromone, and $\nu$ is the rate at which the pheromone dissipates. (See \cite[Chapter 13]{camazine_self-organization_2001} or \cite[Box 3.A]{sumpter_collective_2010} for textbook treatments.) For the simulation shown in \cref{fig: deneubourg}b, we used $\phi = 0.1$, $q = 1$ and $\nu = 1/1500$ \cite{nicolis_emerging_1999}. Since the pheromone is proportional to the number of ants on each path, we plot $x_1/(x_1+x_2) = C_1/(C_1+C_2)$ for equilibrium values of $C_1$ and $C_2$.


\begin{thebibliography}{79}
\providecommand{\natexlab}[1]{#1}
\providecommand{\url}[1]{\texttt{#1}}
\expandafter\ifx\csname urlstyle\endcsname\relax
  \providecommand{\doi}[1]{doi: #1}\else
  \providecommand{\doi}{doi: \begingroup \urlstyle{rm}\Url}\fi

\bibitem[Franks(1989)]{franks_army_1989}
Nigel~R. Franks.
\newblock Army {Ants}: {A} {Collective} {Intelligence}.
\newblock \emph{American Scientist}, 77\penalty0 (2):\penalty0 138--145, 1989.

\bibitem[Anderson(1972)]{anderson_more_1972}
P.~W. Anderson.
\newblock More {Is} {Different}.
\newblock \emph{Science}, 177\penalty0 (4047):\penalty0 393--396, 1972.

\bibitem[Beekman et~al.(2001)Beekman, Sumpter, and Ratnieks]{beekman_phase_2001}
Madeleine Beekman, David J.~T. Sumpter, and Francis L.~W. Ratnieks.
\newblock Phase transition between disordered and ordered foraging in {Pharaoh}'s ants.
\newblock \emph{Proceedings of the National Academy of Sciences}, 98\penalty0 (17):\penalty0 9703--9706, 2001.

\bibitem[Chandra et~al.(2021)Chandra, Gal, and Kronauer]{chandra_colony_2021}
Vikram Chandra, Asaf Gal, and Daniel J.~C. Kronauer.
\newblock Colony expansions underlie the evolution of army ant mass raiding.
\newblock \emph{Proceedings of the National Academy of Sciences}, 118\penalty0 (22):\penalty0 e2026534118, 2021.

\bibitem[Tunstr{\o}m et~al.(2013)Tunstr{\o}m, Katz, Ioannou, Huepe, Lutz, and Couzin]{tunstrom_collective_2013}
Kolbj{\o}rn Tunstr{\o}m, Yael Katz, Christos~C. Ioannou, Cristi{\'a}n Huepe, Matthew~J. Lutz, and Iain~D. Couzin.
\newblock Collective {States}, {Multistability} and {Transitional} {Behavior} in {Schooling} {Fish}.
\newblock \emph{PLOS Computational Biology}, 9\penalty0 (2):\penalty0 e1002915, 2013.

\bibitem[Jhawar et~al.(2020)Jhawar, Morris, Amith-Kumar, Danny~Raj, Rogers, Rajendran, and Guttal]{jhawar_noise-induced_2020}
Jitesh Jhawar, Richard~G. Morris, U.~R. Amith-Kumar, M.~Danny~Raj, Tim Rogers, Harikrishnan Rajendran, and Vishwesha Guttal.
\newblock Noise-induced schooling of fish.
\newblock \emph{Nature Physics}, 16\penalty0 (4):\penalty0 488--493, 2020.

\bibitem[Moberg et~al.(2019)Moberg, Becker, Dierking, Zurheide, Bandow, Buck, Hudait, Molinero, Paesani, and Zeuch]{moberg_end_2019}
Daniel~R. Moberg, Daniel Becker, Christoph~W. Dierking, Florian Zurheide, Bernhard Bandow, Udo Buck, Arpa Hudait, Valeria Molinero, Francesco Paesani, and Thomas Zeuch.
\newblock The end of ice {I}.
\newblock \emph{Proceedings of the National Academy of Sciences}, 116\penalty0 (49):\penalty0 24413--24419, 2019.

\bibitem[Ladyman and Wiesner(2020)]{ladyman_what_2020}
James Ladyman and Karoline Wiesner.
\newblock \emph{What {Is} a {Complex} {System}?}
\newblock Yale University Press, New Haven, 2020.

\bibitem[Romanczuk and Daniels(2022)]{romanczuk_phase_2022}
Pawel Romanczuk and Bryan~C. Daniels.
\newblock Phase {Transitions} and {Criticality} in the {Collective} {Behavior} of {Animals} --- {Self}-{Organization} and {Biological} {Function}.
\newblock In \emph{Order, {Disorder} and {Criticality}}, pages 179--208. World Scientific, 2022.

\bibitem[Saito and Kaneko(2015)]{saito_theoretical_2015}
Nen Saito and Kunihiko Kaneko.
\newblock Theoretical analysis of discreteness-induced transition in autocatalytic reaction dynamics.
\newblock \emph{Phys. Rev. E}, 91\penalty0 (2):\penalty0 022707, 2015.

\bibitem[Togashi and Kaneko(2001)]{togashi_transitions_2001}
Yuichi Togashi and Kunihiko Kaneko.
\newblock Transitions {Induced} by the {Discreteness} of {Molecules} in a {Small} {Autocatalytic} {System}.
\newblock \emph{Physical Review Letters}, 86\penalty0 (11):\penalty0 2459--2462, 2001.

\bibitem[Biancalani et~al.(2014)Biancalani, Dyson, and McKane]{biancalani_noise-induced_2014}
Tommaso Biancalani, Louise Dyson, and Alan~J. McKane.
\newblock Noise-{Induced} {Bistable} {States} and {Their} {Mean} {Switching} {Time} in {Foraging} {Colonies}.
\newblock \emph{Physical Review Letters}, 112\penalty0 (3):\penalty0 038101, 2014.

\bibitem[Van~Kampen(2007)]{van_kampen_chapter_2007}
N.G. Van~Kampen.
\newblock Chapter {V} - {The} {Master} {Equation}.
\newblock In N.G. Van~Kampen, editor, \emph{Stochastic {Processes} in {Physics} and {Chemistry} ({Third} {Edition})}, pages 96--133. Elsevier, Amsterdam, 2007.

\bibitem[Qian et~al.(2002)Qian, Saffarian, and Elson]{qian_concentration_2002}
Hong Qian, Saveez Saffarian, and Elliot~L. Elson.
\newblock Concentration fluctuations in a mesoscopic oscillating chemical reaction system.
\newblock \emph{Proceedings of the National Academy of Sciences}, 99\penalty0 (16):\penalty0 10376--10381, 2002.

\bibitem[Gillespie(2007)]{gillespie_stochastic_2007}
Daniel~T. Gillespie.
\newblock Stochastic {Simulation} of {Chemical} {Kinetics}.
\newblock \emph{Annual Review of Physical Chemistry}, 58\penalty0 (Volume 58, 2007):\penalty0 35--55, 2007.

\bibitem[Gardiner(2009)]{gardiner_stochastic_2009}
Crispin Gardiner.
\newblock \emph{Stochastic {Methods}: {A} {Handbook} for the {Natural} and {Social} {Sciences}}.
\newblock Springer {Series} in {Synergetics}. Springer-Verlag, Berlin Heidelberg, 2009.

\bibitem[Grima et~al.(2011)Grima, Thomas, and Straube]{grima_how_2011}
Ramon Grima, Philipp Thomas, and Arthur~V. Straube.
\newblock How accurate are the nonlinear chemical {Fokker}-{Planck} and chemical {Langevin} equations?
\newblock \emph{The Journal of Chemical Physics}, 135\penalty0 (8):\penalty0 084103, 2011.

\bibitem[Murray(2002)]{murray_mathematical_2002}
J.~D. Murray, editor.
\newblock \emph{Mathematical {Biology}: {I}. {An} {Introduction}}, volume~17 of \emph{Interdisciplinary {Applied} {Mathematics}}.
\newblock Springer, New York, NY, 2002.

\bibitem[Sumpter(2010)]{sumpter_collective_2010}
David J.~T. Sumpter.
\newblock \emph{Collective {Animal} {Behavior}}.
\newblock Princeton University Press, 2010.

\bibitem[Ouellette and Gordon(2021)]{ouellette_goals_2021}
Nicholas~T. Ouellette and Deborah~M. Gordon.
\newblock Goals and {Limitations} of {Modeling} {Collective} {Behavior} in {Biological} {Systems}.
\newblock \emph{Frontiers in Physics}, 9:\penalty0 687823, 2021.

\bibitem[Okubo(1986)]{okubo_dynamical_1986}
Akira Okubo.
\newblock Dynamical aspects of animal grouping: {Swarms}, schools, flocks, and herds.
\newblock \emph{Advances in Biophysics}, 22:\penalty0 1--94, 1986.

\bibitem[Flierl et~al.(1999)Flierl, Gr{\"u}nbaum, Levin, and Olson]{flierl_individuals_1999}
G.~Flierl, D.~Gr{\"u}nbaum, S.~Levin, and D.~Olson.
\newblock From {Individuals} to {Aggregations}: the {Interplay} between {Behavior} and {Physics}.
\newblock \emph{Journal of Theoretical Biology}, 196\penalty0 (4):\penalty0 397--454, 1999.

\bibitem[Camazine et~al.(2001)Camazine, Deneubourg, Franks, Sneyd, Theraulaz, and Bonabeau]{camazine_self-organization_2001}
Scott Camazine, Jean-Louis Deneubourg, Nigel~R. Franks, James Sneyd, Guy Theraulaz, and Eric Bonabeau.
\newblock \emph{Self-{Organization} in {Biological} {Systems}}, volume~38.
\newblock Princeton University Press, 2001.

\bibitem[Buhl et~al.(2006)Buhl, Sumpter, Couzin, Hale, Despland, Miller, and Simpson]{buhl_disorder_2006}
C.~Buhl, D.~J.~T. Sumpter, I.~D. Couzin, J.~J. Hale, E.~Despland, E.~R. Miller, and S.~J. Simpson.
\newblock From {Disorder} to {Order} in {Marching} {Locusts}.
\newblock \emph{Science}, 312\penalty0 (5778):\penalty0 1402--1406, 2006.

\bibitem[Bialek et~al.(2012)Bialek, Cavagna, Giardina, Mora, Silvestri, Viale, and Walczak]{bialek_statistical_2012}
William Bialek, Andrea Cavagna, Irene Giardina, Thierry Mora, Edmondo Silvestri, Massimiliano Viale, and Aleksandra~M. Walczak.
\newblock Statistical mechanics for natural flocks of birds.
\newblock \emph{Proceedings of the National Academy of Sciences}, 109\penalty0 (13):\penalty0 4786--4791, 2012.

\bibitem[Attanasi et~al.(2014)Attanasi, Cavagna, Del~Castello, Giardina, Melillo, Parisi, Pohl, Rossaro, Shen, Silvestri, and Viale]{attanasi_finite-size_2014}
Alessandro Attanasi, Andrea Cavagna, Lorenzo Del~Castello, Irene Giardina, Stefania Melillo, Leonardo Parisi, Oliver Pohl, Bruno Rossaro, Edward Shen, Edmondo Silvestri, and Massimiliano Viale.
\newblock Finite-{Size} {Scaling} as a {Way} to {Probe} {Near}-{Criticality} in {Natural} {Swarms}.
\newblock \emph{Physical Review Letters}, 113\penalty0 (23):\penalty0 238102, 2014.

\bibitem[Ouellette(2022)]{ouellette_physics_2022}
Nicholas~T. Ouellette.
\newblock A physics perspective on collective animal behavior.
\newblock \emph{Physical Biology}, 19\penalty0 (2):\penalty0 021004, 2022.

\bibitem[Brambilla et~al.(2013)Brambilla, Ferrante, Birattari, and Dorigo]{brambilla_swarm_2013}
Manuele Brambilla, Eliseo Ferrante, Mauro Birattari, and Marco Dorigo.
\newblock Swarm robotics: a review from the swarm engineering perspective.
\newblock \emph{Swarm Intelligence}, 7\penalty0 (1):\penalty0 1--41, 2013.

\bibitem[Dorigo et~al.(2014)Dorigo, Birattari, and Brambilla]{dorigo_swarm_2014}
Marco Dorigo, Mauro Birattari, and Manuele Brambilla.
\newblock Swarm robotics.
\newblock \emph{Scholarpedia}, 9\penalty0 (1):\penalty0 1463, 2014.

\bibitem[Hamann(2018)]{hamann_swarm_2018}
Heiko Hamann.
\newblock \emph{Swarm {Robotics}: {A} {Formal} {Approach}}.
\newblock Springer International Publishing, Cham, 2018.

\bibitem[Vicsek et~al.(1995)Vicsek, Czir{\'o}k, Ben-Jacob, Cohen, and Shochet]{vicsek_novel_1995}
Tam{\'a}s Vicsek, Andr{\'a}s Czir{\'o}k, Eshel Ben-Jacob, Inon Cohen, and Ofer Shochet.
\newblock Novel {Type} of {Phase} {Transition} in a {System} of {Self}-{Driven} {Particles}.
\newblock \emph{Physical Review Letters}, 75\penalty0 (6):\penalty0 1226--1229, 1995.

\bibitem[Cates and Tailleur(2015)]{Cates2015}
Michael~E. Cates and Julien Tailleur.
\newblock Motility-induced phase separation.
\newblock \emph{Annual Review of Condensed Matter Physics}, 6\penalty0 (Volume 6, 2015):\penalty0 219--244, 2015.

\bibitem[Horsthemke and Lefever(1984)]{horsthemke_noise-induced_1984}
W.~Horsthemke and R.~Lefever.
\newblock \emph{Noise-{Induced} {Transitions}: {Theory} and {Applications} in {Physics}, {Chemistry}, and {Biology}}.
\newblock Springer {Series} in {Synergetics}. Springer-Verlag, Berlin, 1984.

\bibitem[de~Palma and Lef{\`e}vre(1984)]{de_palma_bifurcation_1984}
Andr{\'e} de~Palma and Claude Lef{\`e}vre.
\newblock Bifurcation and behavior of complex systems.
\newblock \emph{Applied Mathematics and Computation}, 14\penalty0 (1):\penalty0 77--95, 1984.

\bibitem[Vellela and Qian(2008)]{vellela_stochastic_2008}
Melissa Vellela and Hong Qian.
\newblock Stochastic dynamics and non-equilibrium thermodynamics of a bistable chemical system: the {Schl{\"o}gl} model revisited.
\newblock \emph{Journal of The Royal Society Interface}, 6\penalty0 (39):\penalty0 925--940, 2008.

\bibitem[Mendler et~al.(2018)Mendler, Falk, and Drossel]{mendler_analysis_2018}
Marc Mendler, Johannes Falk, and Barbara Drossel.
\newblock Analysis of stochastic bifurcations with phase portraits.
\newblock \emph{PLOS ONE}, 13\penalty0 (4):\penalty0 e0196126, 2018.

\bibitem[Arnold and Boxler(1992)]{arnold_stochastic_1992}
Ludwig Arnold and Petra Boxler.
\newblock Stochastic bifurcation: instructive examples in dimension one.
\newblock In Mark~A. Pinsky and Volker Wihstutz, editors, \emph{Diffusion {Processes} and {Related} {Problems} in {Analysis}, {Volume} {II}: {Stochastic} {Flows}}, Progress in {Probability}, pages 241--255. Birkh{\"a}user, Boston, MA, 1992.

\bibitem[Arnold(1998)]{arnold_random_1998}
Ludwig Arnold.
\newblock \emph{Random {Dynamical} {Systems}}.
\newblock Springer {Monographs} in {Mathematics}. Springer, Berlin, Heidelberg, 1998.

\bibitem[Beckers et~al.(1990)Beckers, Deneubourg, Goss, and Pasteels]{beckers_collective_1990}
R.~Beckers, J.~L. Deneubourg, S.~Goss, and J.~M. Pasteels.
\newblock Collective decision making through food recruitment.
\newblock \emph{Insectes Sociaux}, 37\penalty0 (3):\penalty0 258--267, 1990.

\bibitem[Deneubourg et~al.(1990)Deneubourg, Aron, Goss, and Pasteels]{deneubourg_self-organizing_1990}
J.~L. Deneubourg, S.~Aron, S.~Goss, and J.~M. Pasteels.
\newblock The self-organizing exploratory pattern of the argentine ant.
\newblock \emph{Journal of Insect Behavior}, 3\penalty0 (2):\penalty0 159--168, 1990.

\bibitem[Kirman(1993)]{kirman_ants_1993}
Alan Kirman.
\newblock Ants, {Rationality}, and {Recruitment}.
\newblock \emph{The Quarterly Journal of Economics}, 108\penalty0 (1):\penalty0 137--156, 1993.

\bibitem[Alfarano et~al.(2005)Alfarano, Lux, and Wagner]{alfarano_estimation_2005}
Simone Alfarano, Thomas Lux, and Friedrich Wagner.
\newblock Estimation of {Agent}-{Based} {Models}: {The} {Case} of an {Asymmetric} {Herding} {Model}.
\newblock \emph{Computational Economics}, 26\penalty0 (1):\penalty0 19--49, 2005.

\bibitem[Alfarano and Milakovi{\'c}(2009)]{alfarano_network_2009}
Simone Alfarano and Mishael Milakovi{\'c}.
\newblock Network structure and \textit{{N}}-dependence in agent-based herding models.
\newblock \emph{Journal of Economic Dynamics and Control}, 33\penalty0 (1):\penalty0 78--92, 2009.

\bibitem[Carro et~al.(2015)Carro, Toral, and Miguel]{carro_markets_2015}
Adri{\'a}n Carro, Ra{\'u}l Toral, and Maxi~San Miguel.
\newblock Markets, {Herding} and {Response} to {External} {Information}.
\newblock \emph{PLOS ONE}, 10\penalty0 (7):\penalty0 e0133287, 2015.

\bibitem[Moran et~al.(2020)Moran, Fosset, Benzaquen, and Bouchaud]{moran_schrodingers_2020}
Jos{\'e} Moran, Antoine Fosset, Michael Benzaquen, and Jean-Philippe Bouchaud.
\newblock Schr{\"o}dinger's ants: a continuous description of {Kirman}'s recruitment model.
\newblock \emph{Journal of Physics: Complexity}, 1\penalty0 (3):\penalty0 035002, 2020.

\bibitem[Khalil and Galla(2021)]{khalil_zealots_2021}
Nagi Khalil and Tobias Galla.
\newblock Zealots in multistate noisy voter models.
\newblock \emph{Physical Review E}, 103\penalty0 (1):\penalty0 012311, 2021.

\bibitem[Caligiuri and Galla(2023)]{caligiuri_noisy_2023}
Annalisa Caligiuri and Tobias Galla.
\newblock Noisy voter models in switching environments.
\newblock \emph{Physical Review E}, 108\penalty0 (4):\penalty0 044301, 2023.

\bibitem[Moran(1958)]{moran_random_1958}
P.~A.~P. Moran.
\newblock Random processes in genetics.
\newblock \emph{Mathematical Proceedings of the Cambridge Philosophical Society}, 54\penalty0 (1):\penalty0 60--71, 1958.

\bibitem[Ohkubo et~al.(2008)Ohkubo, Shnerb, and A.~Kessler]{ohkubo_transition_2008}
Jun Ohkubo, Nadav Shnerb, and David A.~Kessler.
\newblock Transition {Phenomena} {Induced} by {Internal} {Noise} and {Quasi}-{Absorbing} {State}.
\newblock \emph{Journal of the Physical Society of Japan}, 77\penalty0 (4):\penalty0 044002, 2008.

\bibitem[McSweeney and Popovic(2014)]{mcsweeney_stochastically-induced_2014}
John~K. McSweeney and Lea Popovic.
\newblock Stochastically-induced bistability in chemical reaction systems.
\newblock \emph{The Annals of Applied Probability}, 24\penalty0 (3):\penalty0 1226--1268, 2014.

\bibitem[Hoessly and Mazza(2019)]{hoessly_stationary_2019}
Linard Hoessly and Christian Mazza.
\newblock Stationary {Distributions} and {Condensation} in {Autocatalytic} {Reaction} {Networks}.
\newblock \emph{SIAM Journal on Applied Mathematics}, 79\penalty0 (4):\penalty0 1173--1196, 2019.

\bibitem[Bibbona et~al.(2020)Bibbona, Kim, and Wiuf]{bibbona_stationary_2020}
Enrico Bibbona, Jinsu Kim, and Carsten Wiuf.
\newblock Stationary distributions of systems with discreteness-induced transitions.
\newblock \emph{Journal of The Royal Society Interface}, 17\penalty0 (168):\penalty0 20200243, 2020.

\bibitem[Gallinger and Popovic(2024)]{gallinger_asymmetric_2024}
Cameron Gallinger and Lea Popovic.
\newblock Asymmetric autocatalytic reactions and their stationary distribution.
\newblock \emph{Royal Society Open Science}, 2024.

\bibitem[Houchmandzadeh and Vallade(2015)]{houchmandzadeh_exact_2015}
Bahram Houchmandzadeh and Marcel Vallade.
\newblock Exact results for a noise-induced bistable system.
\newblock \emph{Physical Review E}, 91\penalty0 (2):\penalty0 022115, 2015.

\bibitem[Holehouse and Moran(2022)]{holehouse_exact_2022}
James Holehouse and Jos{\'e} Moran.
\newblock Exact time-dependent dynamics of discrete binary choice models.
\newblock \emph{Journal of Physics: Complexity}, 3\penalty0 (3):\penalty0 035005, 2022.

\bibitem[Pinsky and Karlin(2011)]{pinsky_introduction_2011}
Mark~A. Pinsky and Samuel Karlin.
\newblock \emph{An {Introduction} to {Stochastic} {Modeling}}.
\newblock Academic Press, Boston, fourth edition, 2011.

\bibitem[Rust et~al.(1995)Rust, Owens, and Reierson]{rust_understanding_1995}
Michael~K Rust, John~M Owens, and Donald~A Reierson, editors.
\newblock \emph{Understanding and {Controlling} the {German} {Cockroach}}.
\newblock Oxford University Press, 1995.

\bibitem[Appel and Smith(1996)]{appel_harborage_1996}
A.~G. Appel and L.~M. Smith, II.
\newblock Harborage {Preferences} of {American} and {Smokybrown} {Cockroaches} ({Dictyoptera}: {Blattidae}) for {Common} {Landscape} {Materials}.
\newblock \emph{Environmental Entomology}, 25\penalty0 (4):\penalty0 817--824, 1996.

\bibitem[Am{\'e} et~al.(2004)Am{\'e}, Rivault, and Deneubourg]{ame_cockroach_2004}
Jean-Marc Am{\'e}, Colette Rivault, and Jean-Louis Deneubourg.
\newblock Cockroach aggregation based on strain odour recognition.
\newblock \emph{Animal Behaviour}, 68\penalty0 (4):\penalty0 793--801, 2004.

\bibitem[Pacala et~al.(1996)Pacala, Gordon, and Godfray]{pacala_effects_1996}
Stephen~W. Pacala, Deborah~M. Gordon, and H.~C.~J. Godfray.
\newblock Effects of social group size on information transfer and task allocation.
\newblock \emph{Evolutionary Ecology}, 10\penalty0 (2):\penalty0 127--165, 1996.

\bibitem[Kurtz(1970)]{kurtz_solutions_1970}
Thomas~G. Kurtz.
\newblock Solutions of ordinary differential equations as limits of pure jump {M}arkov processes.
\newblock \emph{Journal of Applied Probability}, 7\penalty0 (1):\penalty0 49--58, 1970.

\bibitem[Kurtz(1971)]{kurtz_limit_1971}
T.~G. Kurtz.
\newblock Limit theorems for sequences of jump {Markov} processes approximating ordinary differential processes.
\newblock \emph{Journal of Applied Probability}, 8\penalty0 (2):\penalty0 344--356, 1971.

\bibitem[Kurtz(1977)]{kurtz_strong_1977}
Thomas~G. Kurtz.
\newblock Strong approximation theorems for density dependent {Markov} chains.
\newblock \emph{Stochastic Processes and their Applications}, 6\penalty0 (3):\penalty0 223--240, 1977.

\bibitem[Kurtz(1972)]{kurtz_relationship_1972}
Thomas~G. Kurtz.
\newblock The {Relationship} between {Stochastic} and {Deterministic} {Models} for {Chemical} {Reactions}.
\newblock \emph{The Journal of Chemical Physics}, 57\penalty0 (7):\penalty0 2976--2978, 1972.

\bibitem[Keizer(1987)]{keizer_statistical_1987}
Joel Keizer.
\newblock \emph{Statistical {Thermodynamics} of {Nonequilibrium} {Processes}}.
\newblock Springer, New York, NY, 1987.

\bibitem[Vellela and Qian(2007)]{vellela_quasistationary_2007}
Melissa Vellela and Hong Qian.
\newblock A {Quasistationary} {Analysis} of a {Stochastic} {Chemical} {Reaction}: {Keizer}'s {Paradox}.
\newblock \emph{Bulletin of Mathematical Biology}, 69\penalty0 (5):\penalty0 1727--1746, 2007.

\bibitem[Pasteels et~al.(1987)Pasteels, Deneubourg, and Goss]{pasteels_self-organization_1987}
Jacques Pasteels, Jean-Louis Deneubourg, and Simon Goss.
\newblock Self-organization mechanisms in ant societies (i): {Trail} recruitment to newly discovered food sources.
\newblock \emph{From individual to collective behavior in social insects}, pages 155--175, 1987.

\bibitem[Ko et~al.(2022)Ko, Hadgu, Komilian, and Hu]{ko_small_2022}
Hungtang Ko, Mathias Hadgu, Keyana Komilian, and David~L. Hu.
\newblock Small fire ant rafts are unstable.
\newblock \emph{Physical Review Fluids}, 7\penalty0 (9):\penalty0 090501, 2022.

\bibitem[{\c S}ahin(2005)]{sahin_swarm_2005}
Erol {\c S}ahin.
\newblock Swarm {Robotics}: {From} {Sources} of {Inspiration} to {Domains} of {Application}.
\newblock In Erol {\c S}ahin and William~M. Spears, editors, \emph{Swarm {Robotics}}, pages 10--20. Springer Berlin Heidelberg, 2005.

\bibitem[Kelley and Ouellette(2013)]{Kelley2013}
Douglas~H. Kelley and Nicholas~T. Ouellette.
\newblock Emergent dynamics of laboratory insect swarms.
\newblock \emph{Scientific Reports}, 3\penalty0 (1):\penalty0 1073, 2013.

\bibitem[Puckett and Ouellette(2014)]{puckett_determining_2014}
James~G. Puckett and Nicholas~T. Ouellette.
\newblock Determining asymptotically large population sizes in insect swarms.
\newblock \emph{Journal of The Royal Society Interface}, 11\penalty0 (99):\penalty0 20140710, 2014.

\bibitem[Gordon et~al.(1993)Gordon, Paul, and Thorpe]{Gordon1993}
Deborah~M. Gordon, Richard~E. Paul, and Karen Thorpe.
\newblock What is the function of encounter patterns in ant colonies?
\newblock \emph{Animal Behaviour}, 45\penalty0 (6):\penalty0 1083--1100, 1993.

\bibitem[Anderson et~al.(2010)Anderson, Craciun, and Kurtz]{anderson_product-form_2010}
David~F. Anderson, Gheorghe Craciun, and Thomas~G. Kurtz.
\newblock Product-{Form} {Stationary} {Distributions} for {Deficiency} {Zero} {Chemical} {Reaction} {Networks}.
\newblock \emph{Bulletin of Mathematical Biology}, 72\penalty0 (8):\penalty0 1947--1970, 2010.

\bibitem[Mendler and Drossel(2020)]{mendler_predicting_2020}
Marc Mendler and Barbara Drossel.
\newblock Predicting properties of the stationary probability currents for two-species reaction systems without solving the {Fokker}-{Planck} equation.
\newblock \emph{Physical Review E}, 102\penalty0 (2):\penalty0 022208, 2020.

\bibitem[Becker et~al.(2020)Becker, Mendler, and Drossel]{becker_relation_2020}
Lara Becker, Marc Mendler, and Barbara Drossel.
\newblock Relation between the convective field and the stationary probability distribution of chemical reaction networks.
\newblock \emph{New Journal of Physics}, 22\penalty0 (3):\penalty0 033012, 2020.

\bibitem[Dyson et~al.(2015)Dyson, Yates, Buhl, and McKane]{PhysRevE.92.052708}
Louise Dyson, Christian~A. Yates, Camille Buhl, and Alan~J. McKane.
\newblock Onset of collective motion in locusts is captured by a minimal model.
\newblock \emph{Physical Review E}, 92:\penalty0 052708, 2015.

\bibitem[Nicolis and Deneubourg(1999)]{nicolis_emerging_1999}
S.~C Nicolis and J.~L Deneubourg.
\newblock Emerging {Patterns} and {Food} {Recruitment} in {Ants}: an {Analytical} {Study}.
\newblock \emph{Journal of Theoretical Biology}, 198\penalty0 (4):\penalty0 575--592, 1999.

\bibitem[Beckers et~al.(1992)Beckers, Deneubourg, and Goss]{beckers_trails_1992}
R.~Beckers, J.~L. Deneubourg, and S.~Goss.
\newblock Trails and {U}-turns in the selection of a path by the ant \textit{{Lasius} niger}.
\newblock \emph{Journal of Theoretical Biology}, 159\penalty0 (4):\penalty0 397--415, 1992.

\bibitem[Beckers et~al.(1993)Beckers, Deneubourg, and Goss]{beckers_modulation_1993}
R.~Beckers, J.~L. Deneubourg, and S.~Goss.
\newblock Modulation of trail laying in the ant {Lasius} niger ({Hymenoptera}: {Formicidae}) and its role in the collective selection of a food source.
\newblock \emph{Journal of Insect Behavior}, 6\penalty0 (6):\penalty0 751--759, 1993.

\end{thebibliography}

\end{document}